\documentclass[conference]{IEEEtran}
\IEEEoverridecommandlockouts
\def\BibTeX{{\rm B\kern-.05em{\sc i\kern-.025em b}\kern-.08em
    T\kern-.1667em\lower.7ex\hbox{E}\kern-.125emX}}

\usepackage{indentfirst}
\usepackage{graphicx}
\usepackage{amssymb}
\usepackage{amsfonts}
\usepackage{mathrsfs}
\usepackage{leftidx}
\usepackage{color}
\usepackage{amsmath}
\usepackage{arydshln}
\usepackage{ragged2e}
\usepackage{cite}
\usepackage{enumerate}
\usepackage{longtable}
\usepackage{float}
\usepackage{stfloats}
\usepackage{algpseudocode}
\usepackage{algorithm}
\usepackage[caption=false,font=normalsize,labelfont=sf,textfont=sf]{subfig}
\usepackage{tabularx}

\usepackage[table,usenames,dvipsnames]{xcolor}
\definecolor{aa}{RGB}{175,238,238}
\definecolor{bb}{RGB}{255,255,255}

\usepackage{bm}
\usepackage{makecell}

\begin{document}

\title{TDGCN-Based Mobile Multiuser Physical-Layer Authentication for EI-Enabled IIoT}









\author{
\IEEEauthorblockN{Rui Meng\IEEEauthorrefmark{1}, Hangyu Zhao\IEEEauthorrefmark{2}, Liang Jin\IEEEauthorrefmark{1}, Bingxuan Xu\IEEEauthorrefmark{1}, Ce Liu\IEEEauthorrefmark{1}, and Xiaodong Xu\IEEEauthorrefmark{1}}
\IEEEauthorblockA{\IEEEauthorrefmark{1} State Key Laboratory of Networking and Switching Technology, BUPT, Beijing, China}
\IEEEauthorblockA{\IEEEauthorrefmark{2} Chinese University of Hong Kong, Hong Kong SAR, China}
\IEEEauthorblockA{\{buptmengrui, jinliang, xubingxuan, liuce, xuxiaodong\}@bupt.edu.cn, zhaohangyu@link.cuhk.edu.hk}

\thanks{This work was supported in part by the National Key Research and Development Program of China under Grant 2020YFB1806905; in part by the National Natural Science Foundation of China under Grant 62501066 and Grant U24B20131; and in part by the Beijing Natural Science Foundation under Grant L242012.}
}

\maketitle

\begin{abstract}
Physical-Layer Authentication (PLA) offers endogenous security, lightweight implementation, and high reliability, making it a promising complement to upper-layer security methods in Edge Intelligence (EI)-empowered Industrial Internet of Things (IIoT). However, state-of-the-art Channel State Information (CSI)-based PLA schemes face challenges in recognizing mobile multi-users due to the constantly shifting CSI distributions with user movements. To address this issue, we propose a Temporal Dynamic Graph Convolutional Network (TDGCN)-based PLA scheme, which employs Graph Neural Networks (GNNs) to capture the spatio-temporal dynamics induced by user movements. Firstly, we partition CSI fingerprints into multivariate time series and utilize dynamic GNNs to capture their associations. Secondly, Temporal Convolutional Networks (TCNs) handle temporal dependencies within each CSI fingerprint dimension. Additionally, Dynamic Graph Isomorphism Networks (GINs) and cascade node clustering pooling further enable efficient information aggregation and reduced computational complexity. Simulations demonstrate the proposed scheme's superior authentication accuracy compared to seven baseline schemes.
\end{abstract}

\begin{IEEEkeywords}
Physical-Layer Authentication (PLA), mobile multiuser authentication, IIoT.
\end{IEEEkeywords}

\section{Introduction}

Industrial Internet of Things (IIoT) can enable data collection, analysis, and sharing, ultimately optimizing production processes, monitoring device status, as well as predicting maintenance needs. Edge Intelligence (EI)-enabled IIoT further integrates Artificial Intelligence (AI) into edge networks to reduce latency and bandwidth requirements, improve real-time decision-making capability, and enhance data privacy and security \cite{gu2024ai,meng2025semantic}. However, the inherent openness of radio channels makes these communications susceptible to potential eavesdropping, interception, and forgery attacks.



Channel State Information (CSI)-based Physical-Layer Authentication (PLA) presents a promising supplementary approach to traditional upper-layer authentication methods \cite{xie2020survey,meng2025survey,meng2025generative}. By leveraging the distinct random characteristics of communication links, PLA can provide inherently secure identity protection for transmitters \cite{gao2024esanet}. This becomes particularly advantageous in EI-empowered IIoT, because edge servers are physically proximate to end devices and can conveniently acquire physical-layer attributes \cite{xie2021weighted}.
In earlier literature, PLA is formulated as a statistical hypothesis test, in which the detection threshold is established to identify whether the signal is legal or not \cite{xiao2008using}. Since it is challenging to distinguish multi-users by establishing multi-thresholds, researchers have recently formulated the multiuser PLA as a multi-classification problem and solved it via AI techniques \cite{pan2019threshold,liao2019multiuser,meng2023multiuser,chen2022physical}.

However, existing CSI-based PLA schemes still face challenges in authenticating multiple mobile industrial devices. Most schemes \cite{meng2024multiobservation,meng2023multiuser,liao2019multiuser,liao2019security,chen2022physical} assume that CSI fingerprint of each transmitter follows an independent and identically distributed pattern.
Meng et al. \cite{meng2023multiuser} demonstrate that, the performance of CSI-based schemes decreases as transmitters move away. This is due to CSI is location-specific, and user movement leads to changes in the distribution of CSI. Although researchers have proposed the model-driven scheme \cite{han2024model}, the Long Short-Term Memory (LSTM)-based scheme \cite{germain2021mobile}, the channel prediction-based scheme \cite{wang2021channel}, and the knowledge-enhanced scheme \cite{wang2024knowledge} to improve the performance of PLA in mobile conditions, they are difficult to fully learn latent dependency relationships between each CSI dimension.

Against this background, we analyze CSI fingerprints of mobile users as time series data and employ graph convolutional networks (GCNs) to grasp the evolving patterns of fingerprints across time and space. Since the similarity of fingerprints decreases with users' distance, we model the fingerprints of mobile users as time series data to depict the dynamic changes of fingerprints during user movement. These sequences of fingerprint samples are then represented as graphs, where nodes denote each fingerprint sequence and edges signify the connections or interactions among them. GCNs excel at learning the topological structure and connections between nodes, making them ideal for handling intricate nonlinear relationships and assimilating local and global information among nodes. Consequently, GCNs can effectively capture both the spatio-temporal dynamics within individual time series and the interactions between them, thereby facilitating the authentication of mobile users' identities. The main contributions are summarized as follows.
\begin{enumerate}
    \item We propose the Temporal Dynamic Graph Convolutional Network (TDGCN)-based PLA scheme to achieve reliable mobile multiuser authentication in EI-enabled IIoT.
    \item CSI fingerprints are modeled as time series data, with dynamic GNNs capturing associations between them. Unlike CNNs, GNNs consider latent strong dependencies between each CSI dimension. Within each dynamic GNN, nodes and edges respectively represent CSI sequences and their interactions, with connections further represented by learnable adjacency matrices.
    \item Temporal Convolutional Networks (TCNs) capture temporal dependencies within each CSI dimension. The length of learned features is synchronized with that of CSI sequences through padding operations before input into dynamic GNNs.
    \item Dynamic Graph Isomorphism Networks (GINs) determine whether two graphs are structurally identical or isomorphic, aggregating information in parallel. Cascade node clustering pooling preserves learned information and reduces computational complexity. 
    \item Simulations on synthetic data demonstrate the superior authentication accuracy of the proposed TDGCN-based PLA over seven typical ML-based PLA schemes.
\end{enumerate}

\section{System Model and Problem Formulation}
\subsection{Network Model}


We consider a typical Alice-Eve-Bob model, and the nodes involved are described as follows.

\textit{Alices:} $K_A$ legal industrial terminals are in communication with the legitimate receiver (Bob) during different time slots. Alices are continuously moving to facilitate date collection or meet production line demands, thereby enhancing flexibility and enabling real-time analysis. The distance between transmitters is assumed to be greater than half a wavelength to ensure the distinguishability of their fingerprints\cite{xie2020survey,meng2024multiobservation}.

\textit{Eves:}
$K_E$ spoofing attackers attempt to impersonate the identity information of Alices, such as medium access control (MAC) addresses, to establish communication with Bob \cite{meng2023physical}. 


\textit{Bob:} Bob is positioned at the edge of IIoT to conveniently collect fingerprint samples of terminals, and is equipped with edge servers that offer ample computing power to train the authentication model \cite{xie2021weighted}. Bob aims to identify the transmitter of the received signal using the trained PLA model.

\subsection{Channel Model}
The received signal at Bob is represented as $\bm{b}_s=\bm{x}\bm{a}_s+\bm{n}$, where $\bm{a}_s$ denotes the signal transmitted from Alices/Eves and $\bm{n}\sim\mathcal{CN}(0,\bm{\sigma}^2)$ represents the Gaussian noises. $\bm{x}$ denotes the channel matrix from Alices/Eves to Bob
CSI fingerprints $\bm{x}$ can be acquired through channel estimation.

\subsection{Problem Formulation}
We consider a multiple-input multiple-output (MIMO) scene, and let $N_T$ and $N_R$ respectively denote the number of antennas of Alices/Eves and Bob. The hierarchical CSI fingerprints are multidimensional matrices associated with the positions of devices. Therefore, in moving scenarios, CSI fingerprints can be modeled as multivariate time series (MTS) $\bm{X}=\{\bm{x}_1, \bm{x}_2, ..., \bm{x}_d\}\in\mathbb{R}^{d\times l}$, where $d=2N_RN_T$ denotes the dimension of CSI fingerprints and $l\in\mathbb{N}^*$ represents the length of CSI fingerprint series. $\bm{x}_i=\{x_{i,1},x_{i,2},...,x_{i,l}\}$ ($i\in[1,d]$) represents the sequence of the $i$-th dimension feature in the multi-dimensional CSI fingerprint. 
The mobile multiuser PLA problem involves formulating a classifier $f(\cdot)$ from $\chi=\{\bm{X}_{1}, \bm{X}_{2}, ..., \bm{X}_{M}\}$ to $\eta=\{\bm{y}_1,\bm{y}_2,...,\bm{y}_M\}$ to predict the identity label $\bm{y}_m$ corresponding to the CSI fingerprint sequence $\bm{X}_m$ ($m=[1,M]$).

\section{TDGCN-Based Mobile Multiuser PLA}

\begin{figure*}
\centering
\includegraphics[width=0.9\textwidth]{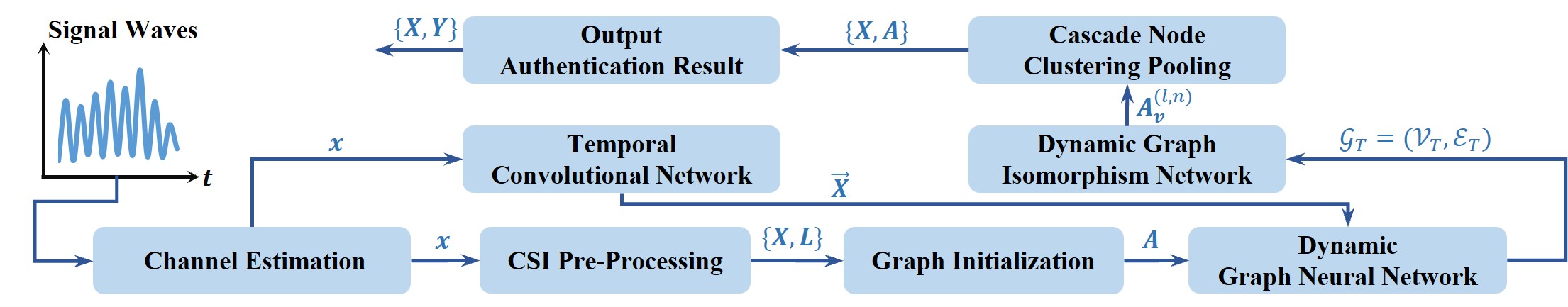}
\caption{Proposed TDGCN-based PLA scheme. 
}
\label{scheme}
\end{figure*}

As depicted in Fig. \ref{scheme}, the proposed TDGCN-based PLA scheme comprises several key modules, with descriptions as follows.

\subsection{CSI Pre-Processing Module}
The training dataset $\mathcal{D}_{\mathrm{train}}$ and testing dataset $\mathcal{D}_{\mathrm{test}}$ are used to train the mobile multiuser authentication model and verify its authentication performance. $\mathcal{D}_{\mathrm{train}}$ is composed of CSI fingerprint sequences $\bm{X}_{\mathrm{train}}$ and corresponding labels $\bm{Y}_{\mathrm{train}}$, which are respectively denoted by
\begin{equation}
\bm{X}_{\mathrm{train}}=[\underbrace{\bm{X}_1^{1},...,\bm{X}_1^{N_1}}_{N_1},\underbrace{\bm{X}_2^{1},...,\bm{X}_2^{N_2}}_{N_2},...,\underbrace{\bm{X}_K^{1},...,\bm{X}_K^{N_K}}_{N_K}]
\end{equation}
and
\begin{equation}
\bm{Y}_{\mathrm{train}}=[\underbrace{\bm{L}_1,...,\bm{L}_1}_{N_1},\underbrace{\bm{L}_2,...,\bm{L}_2}_{N_2},...,\underbrace{\bm{L}_K,...,\bm{L}_K}_{N_K}],
\end{equation}
where $N_k$ is the number of CSI fingerprint sequences of the $k$-th ($k\in[1,K]$) transmitter and $K=K_A+K_E$ is the number of transmitters. $\bm{L}_k$ is the identity label of the $k$-th transmitter, represented by one-hot coding 
as
$\bm{L}_k=[0,...,1,...,0]^T$,
where the $k$-th element is 1 and the others are 0. The CSI sequences of each transmitter are evenly divided by some equidistant time slots $T=\{T_1,T_2,...,T_N\}$ arranged in time sequence, where $N=N_1=...=N_K$ is the number of time slots.

\subsection{Graph Initialization Module}

CNNs have been extensively employed to capture the spatial-frequency features of multidimensional CSI fingerprints in MIMO systems, as seen in \cite{liao2019security,meng2023multiuser,gao2024esanet}. However, existing CNN-based PLA models often overlook the latent dependency relationships between each CSI dimension.
Recognizing that the dependency relationships can be naturally represented as graphs, we propose a graph-based approach to address this challenge.   
The fundamental structure of a graph consists of nodes and edges, commonly denoted as $\mathcal{G}=(\mathcal{V},\mathcal{E})$. Nodes $\mathcal{V}$ represent CSI fingerprint sequences and are the basic building blocks of the graph $\mathcal{G}$. Edges $\mathcal{E}$ serve as connectors between nodes $\mathcal{V}$, revealing the relationships and interactions among them. Edges $\mathcal{E}$ can be either directed or undirected, and they can be assigned weights to quantify the strength or significance of connections between nodes $\mathcal{V}$. Compared with traditional graph structures, GNNs can produce more enriched and insightful node representations leveraging DL-based node learning and updating continually. 

The latent relationships between CSI sequences $\bm{X}$ are modeled by the adjacency matrix. Firstly, the similarity matrix $\bm{S}$ between each dimension fingerprint $\bm{x}$ is calculated by
\begin{equation}
\bm{S}_{ij}=\frac{\mathrm{exp}\left(-\sigma_{\mathrm{ReLU}}(d_{\mathrm{Eu}}(\bm{x}_i,\bm{x}_j))\right)}{\sum_{m=1}^d\mathrm{exp}\left(-\sigma_{\mathrm{ReLU}}(d_{\mathrm{Eu}}(\bm{x}_i,\bm{x}_m)\right)},
\end{equation}
where $\sigma_{\mathrm{ReLU}}(x)=\mathrm{max}(0,x)$ denotes the Rectified Linear Unit (ReLU) activation function and $d_{\mathrm{Eu}}(\bm{x}_i,\bm{x}_j)$ represents the Euclidean distance between $\bm{x}_i$ and $\bm{x}_j$. Then, the adjacency matrix $\bm{A}$ is obtained by 
$\bm{A}=\sigma_{\mathrm{ReLU}}(\bm{S}\bm{\Upsilon})$,
where $\bm{\Upsilon}$ represents learnable parameters.
Moreover, $\bm{A}$ undergoes a sparsification process, wherein a significant portion of its elements is set to 0, rendering the matrix sparser and reducing the computational load. Specifically, the threshold $\theta$ is introduced for normalization as 
$A_{ij}=\{\begin{matrix}A_{ij},&A_{ij}\ge \theta\\\quad0,&A_{ij}<\theta\end{matrix}$.

\subsection{Temporal Convolutional Network Module}
The Temporal Convolutional Network (TCN) module is designed to capture the temporal dependencies between {$x_{i,1}$, $x_{i,2}$, ..., and $x_{i,l}$. It integrates multiple convolution layers with distinct kernels to grasp local characteristics. The extracted features in the $l$-th CNN layer are represented as
\begin{equation}
\bm{Z}_l=\sigma_{\mathrm{ReLU}}(\bm{W}_l * \bm{Z}_{l-1}+\bm{B}_l),
\end{equation}
where $\bm{Z}_{l-1}$ is both the output of the $(l-1)$-th CNN layer and the input of the $l$-th CNN layer, $*$ is the convolution operation, and $\bm{W}_l$ and $\bm{B}_l$ are the weight and bias matrices in the $l$-th CNN layer, respectively.
Causal convolution is employed to guarantee the forward propagation of information during convolution operations \cite{sun2022toward}. Padding operations are subsequently employed to synchronize the length of the output features $\Vec{\bm{X}}$ with that of CSI fingerprint sequences $\bm{X}$.

\subsection{Dynamic Graph Neural Network Module}

GNNs are categorized into static and dynamic graphs. Static graphs are ideal for unchanged topological structures, such as user relationship graphs in social networks, while dynamic graphs excel at managing evolving graph structures and attributes, like traffic networks where vehicle positions change over time. In mobile scenarios, shifts in user positions lead to continual changes in CSI fingerprint distribution. Consequently, dynamic graphs are utilized to capture the temporal dynamics of CSI fingerprint sequences.
For a set of fixed nodes $\mathcal{V}$, the dynamic graph is usually represented as 
\begin{equation}
\label{tra_graph}
    \mathcal{G}_T=(\mathcal{V},\mathcal{E}_T),
\end{equation}
where $\mathcal{G}_T=\{\mathcal{G}_{T_1},...,\mathcal{G}_{T_N}\}$ and $\mathcal{E}_T=\{\mathcal{E}_{T_1}, ..., \mathcal{E}_{T_N}\}$. However, as transmitters move, the nodes $\mathcal{V}$ representing fingerprint sequences $\bm{X}$ are no longer fixed, but vary with channel environments. Therefore, (\ref{tra_graph}) is inappropriate, and we instead introduce the dynamic graph as
$\mathcal{G}_T=(\mathcal{V}_T,\mathcal{E}_T)$.
We assume that the CSI fingerprint sequence evolve from its earlier time slots through information aggregation. As depicted in Fig. \ref{dynamicgraph}, except for the graph $\mathcal{G}_{T_1}$ corresponding to the first fingerprint sequence $\bm{X}^1$, $l$ nodes are introduced to each subsequent graph $\mathcal{G}_{n}$ to represent the node characteristics $\mathcal{V}^{n-1}$ in the graph $\mathcal{G}_{n-1}$ corresponding to the preceding fingerprint sequence $\bm{X}^{n-1}$. Directed edges are then established between the nodes $\mathcal{V}^{n-1}$ of the previous fingerprint sequence $\bm{X}^{n-1}$ and the nodes $\mathcal{V}^{n-1}$ of the current fingerprint sequence $\bm{X}^{n}$ to represent associations. These new directed edges aggregate the nodes $\mathcal{V}^{n-1}$ from the previous graph $\mathcal{G}_{n-1}$ into the nodes $\mathcal{V}^{n}$ of the current graph $\mathcal{G}_{n}$, after which the source nodes $\mathcal{V}^{n-1}$ are removed to maintain consistent node counts across all graphs \{$\mathcal{G}_{2},...,\mathcal{G}_{N}$\} relative to the first graph $\mathcal{G}_{1}$.

\begin{figure}
\centering
\includegraphics[width=0.4\textwidth]{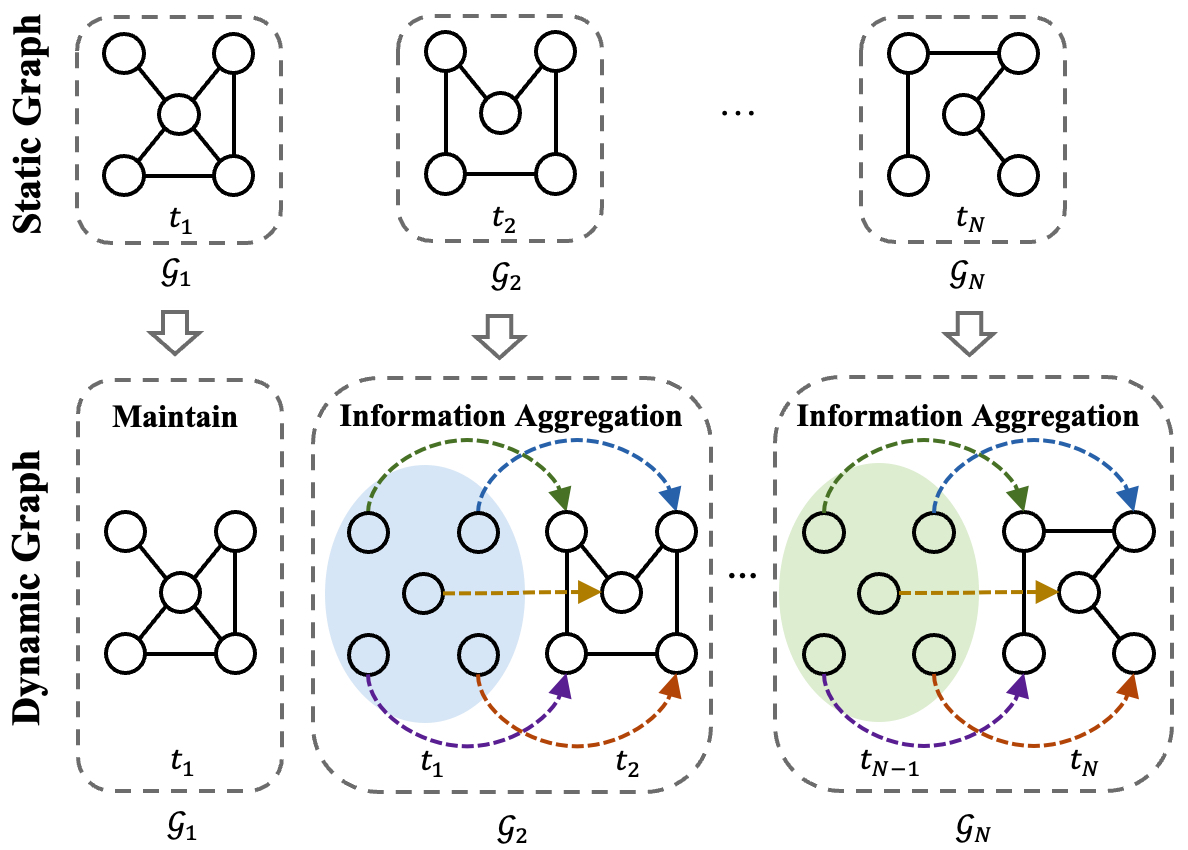}
\caption{Dynamic Graph Transformation.}
\label{dynamicgraph}
\end{figure}

\subsection{Dynamic Graph Isomorphism Network Module}
GINs are hailed as leading variants of GNNs, boasting discriminative and representational prowess comparable to the Weisfeiler-Lehman (WL) graph isomorphism test \cite{NEURIPS2019_bb04af0f}. GINs update node representations as
\begin{equation}
\label{gin}
\bm{A}_v^{l+1}=\mathrm{MLP}^l\left((1+\bm{\varepsilon}^l)\bm{A}_v^l+\sum_{u\in\mathcal{N}(v)}\bm{A}_u^l\right),
\end{equation}
where $\bm{A}^l_v$ is the adjacency matrix of the $v$-th node in the $l$-th layer, $\bm{\varepsilon}^l$ denotes learnable parameters, and $\mathrm{MLP}$ represents Multilayer Perceptron. In contrast to traditional GNNs, GINs replace the mean aggregator with a sum aggregator for nodes and ensure each neighbor contributes equally to updating the central node. Additionally, GINs amalgamate information from all layers of nodes to derive the final representation as
\begin{equation}
\label{gin2}
\bm{A}=\mathrm{CONCAT}\left(\sum_{k=0}^L {\bm{A}_v^k}\right),
\end{equation}
where $\mathrm{CONCAT}$ is the concatenate function.
Due to dynamic CSI sequences, (\ref{gin}) and (\ref{gin2}) are not suitable for dynamic GNNs. Motivated by \cite{liu2023todynet}, Dynamic GINs are employed to aggregate information from different sets of nodes as

\begin{equation}
\begin{aligned}
\bm{A}_{v}^{(l,n)}=& \mathrm{MLP}^{(l,n)}\Big((1+\bm{\epsilon}^{l})\cdot \bm{A}_{v}^{(l-1,n)}+\bm{A}_{v}^{(l-1,n-1)}+  \\
&\sum_{u\in\mathcal{N}(v)}\tilde{\bm{\omega}}_{ij}\cdot \bm{A}_{u}^{(l-1,n)}\Big)
\end{aligned}  
\end{equation}
and

\begin{equation}
\bm{A}_v^{l}=\mathrm{CONCAT}\left(\sum_{n=1}^N\bm{A}_v^{(l,n)}\right),
\end{equation}
where $\bm{A}_{v}^{(l,n)}$ denotes the adjacency matrix for the $v$-th node at the $n$-th time slot in the $l$-th layer, and $\tilde{\bm{\omega}}_{ij}$ represents the normalized weights of edges.

\subsection{Cascade Node Clustering Pooling Module}
Graph pooling is a pivotal component of GNNs, which is similar to the role of pooling operations in traditional neural networks. Its purpose is to diminish the graph's scale, decrease computational complexity, and distill crucial graph features. By aggregating nodes or subgraphs into higher-level representations, graph pooling operations enhance the model's comprehension of the graph's structure and content, thereby boosting its generalization ability \cite{liu2023graph}.

\begin{figure*}
\centering
\includegraphics[width=0.85\textwidth]{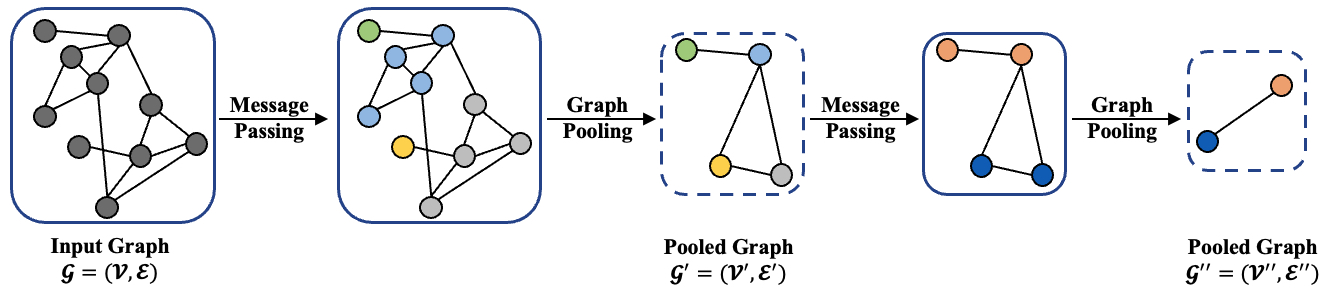}
\caption{Cascade node clustering pooling.}
\label{pooling}
\end{figure*}

As illustrated in Fig. \ref{pooling}, the cascade node clustering pooling module views graph pooling as a node clustering problem, where nodes are mapped into clusters, forming new nodes for the coarsened graph. The cluster assignment matrices predict node assignments in the $l$-th layer as
$\bm{C}^{l}=f_{\mathrm{CA}}(\bm{X}^l,\bm{A}^l)$,
where $f_{\mathrm{CA}}$ denotes the cluster assignment function. Subsequently, new graphs with fewer nodes then are obtained as
\begin{equation}
\{\bm{X}^{l+1},\bm{A}^{l+1}\}=f_{\mathrm{GC}}(\bm{X}^l,\bm{A}^l,\bm{C}^l),
\end{equation}
where $f_{\mathrm{GC}}$ symbolizes the graph coarsening function.

\subsection{Authentication Result Output Module}
The module employs average pooling to compute the average of graph features, yielding a fixed-length vector. Subsequently, this vector is mapped to a logical vector via a fully connected layer, culminating in the authentication result being derived through the softmax function.
The loss function is given as
$\mathcal{L}=-\frac{1}{N}\sum_{i=1}^{N\cdot K}\sum_{j=1}^{K}\bm{y}_{ij}\mathrm{log}\hat{\bm{y}}_{ij}$,
where $\bm{y}_{ij}$ and $\hat{\bm{y}}_{ij}$ are real and predicted identity labels, respectively.
\section{Simulation Results and Analysis}

\textit{Performance Metrics:}
False alarm rate and miss detection rate are typically used to gauge the reliability of PLA models. However, these coarse-grained metrics might not be suitable for multiuser scenarios as they overlook which legal or illegal transmitter the received signal originates from. Thus, drawing inspiration from \cite{meng2023multiuser,meng2024multiobservation,jing2023multi,liao2019multiuser}, the reliability of the proposed multiuser PLA model is evaluated by authentication accuracy, defined as
$P_{\mathrm{accuracy}}=\frac{1}{N\cdot K}\sum_{n=1}^{N \cdot K}\mathbb{I}(\bm{L}_{n}=\bm{Y}_{n})$,
where $N\cdot K$ is the number of CSI fingerprint sequences, $\bm{L}_{n}$ and $\bm{Y}_{n}$ represent actual and predicted identity labels of the $n$-th CSI fingerprint sequences, respectively. $\mathbb{I}$ is the indicator function, defined as $\mathbb{I}(\cdot)= \{\begin{matrix}1,& \cdot \text{ is true}\\0,&\cdot \text{ is false}\end{matrix}$.

\begin{figure}
\centering
\includegraphics[width=0.4\textwidth]{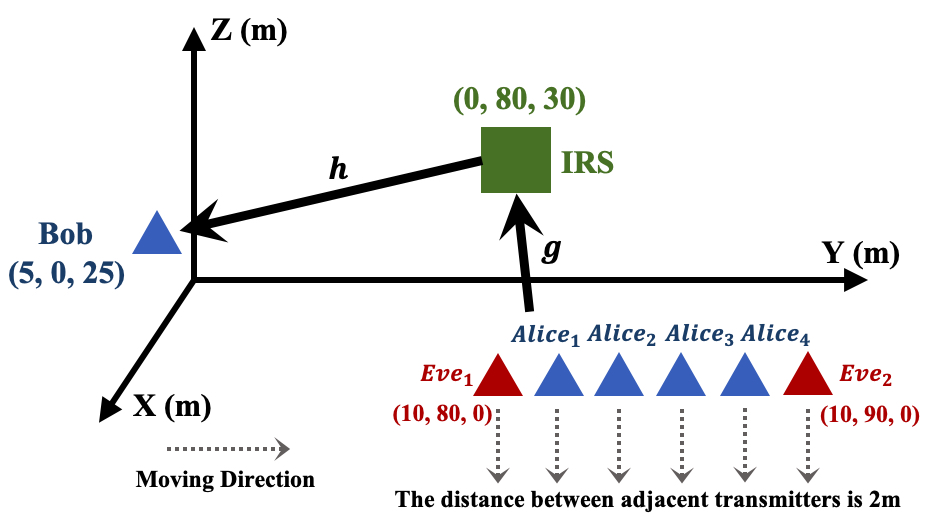}
\caption{Positions of Alices, Eves, IRSs and Bob, where the positions of Alices are (10, 82/84/86/88, 0)}
\label{position}
\end{figure}

\textit{Simulation Parameters:}
As illustrated in Fig. \ref{position}, four legal transmitters and two spoofing attackers are considered. CSI fingerprints are generated through MatLab. We introduce intelligent reconfigurable surfaces to enhance the accuracy and reliability of CSI fingerprints.
The path loss is modeled according to 3GPP TR 38.901. The LoS paths are modeled according to \cite{hu2021reconfigurable}, while the NLoS paths are modeled as Rayleigh fading models. The simulation parameters are given in Tab. \ref{parameter} in detail. The computer configurations are Intel Core i5-13600KF, 3.50 GHz basic frequency, and 32 GB of RAM.

\begin{table}[t]
	\caption {Simulation And Hyper Parameters}
	\label{parameter}
 \begin{center}
\begin{tabular}{|c| c|} 
\hline
\textbf{Parameters}                    & \textbf{Values} \\ \hline
Number of antennas of each transmitter $N_T$ & 4               \\ \hline
Number of antennas of Bob $N_R$              & 3               \\ \hline
Number of IRS elements                 & 8*16            \\ \hline
Carrier Frequency                & 3.5 GHz             \\ \hline
Bandwidth                       & 1 MHz              \\ \hline
Speed of Transmitters & 2 m/s \\ \hline
CSI sampling frequency & 100 Hz \\ \hline
Number of each transmitter's CSI samples    & 50000 \\ \hline
Number of each transmitter's CSI sequences  & 1000  \\ \hline
Length of each CSI sequence & 50 \\ \hline
Number of each transmitter's training CSI samples & 30000 \\ \hline
Number of each transmitter's testing CSI samples & 20000 \\ \hline
Learning rate & 0.0001 \\ \hline
Batch size & 16 \\ \hline
The number of GNN layers & 3 \\ \hline
Time convolutional kernel size for each layer & 9, 5, and 3 \\ \hline
The ratio of pooling for nodes & 0.2 \\ \hline
Decrease rate of weights & 0.0001 \\ \hline
Seed for initializing training & 42 \\ \hline
\end{tabular}
 \end{center}
\end{table}

\textit{Baseline Models:}
Seven PLA models are compared, including Decision Tree (DT)\cite{pan2019threshold}, \textit{K}-Nearest Neighbor (KNN)\cite{pan2019threshold}, Naive Bayesian (NB)\cite{denis2020device}, Weighted Voting (WV)\cite{xie2021weighted}, Gradient Boosting Decision Tree (GBDT)\cite{Douiba2023}, Regularized Gradient Boosting Optimization (RGBO)\cite{meng2024multiobservation}, and Improved Gradient Boosting Optimization (IGBO)\cite{meng2024multiobservation}.





\begin{figure}
\centering
\includegraphics[width=0.35\textwidth]{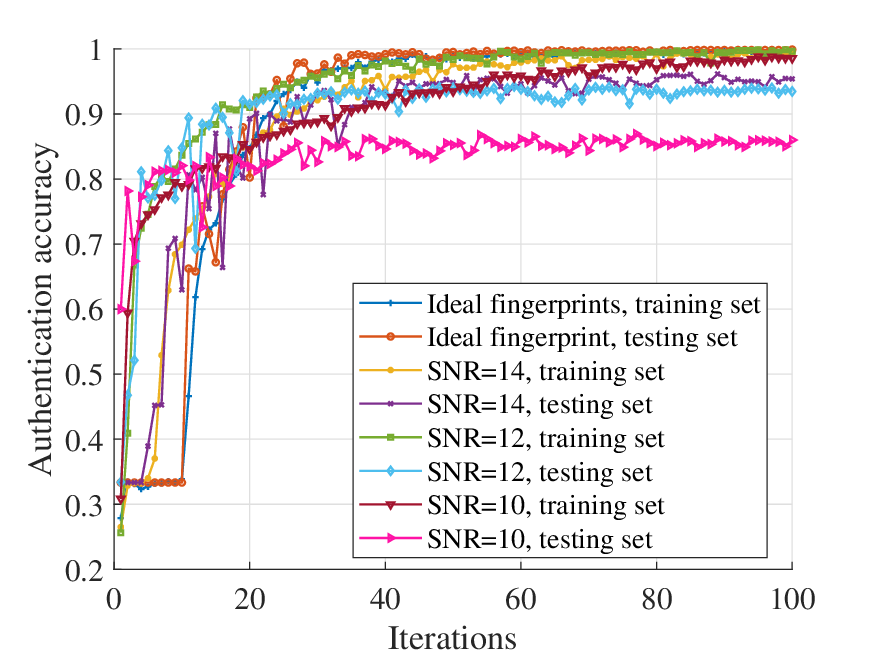}
\caption{Authentication accuracy of the proposed TDGCN-based PLA scheme versus different iteration numbers under different SNRs.}
\label{SNR1}
\end{figure}

\begin{figure}
\centering
\includegraphics[width=0.35\textwidth]{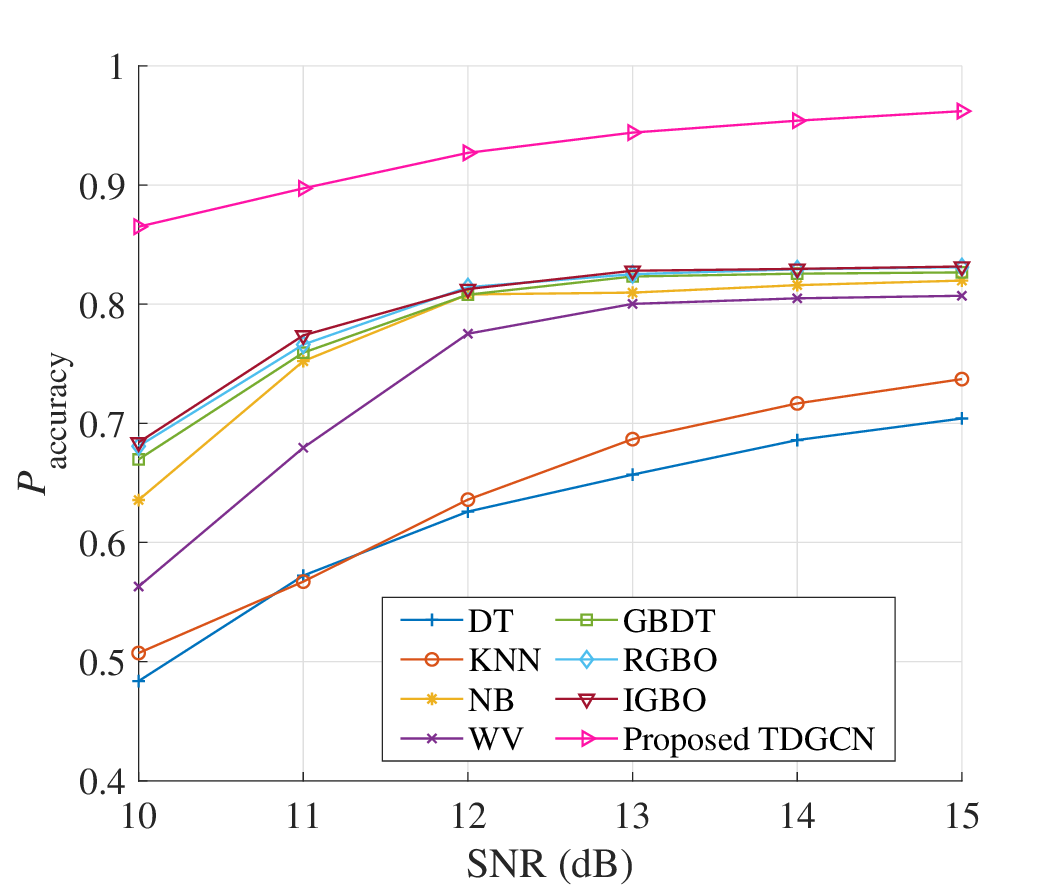}
\caption{Authentication accuracy of different PLA schemes versus SNRs.}
\label{SNR2}
\end{figure}

\textit{Performance under Different SNRs:}
Fig. \ref{SNR1} and Fig. \ref{SNR2} showcase the authentication accuracy across varying SNRs, with artificial noise added to simulate noisy environments. Under ideal CSI conditions, the proposed scheme achieves 100\% authentication accuracy. As SNRs decrease, authentication accuracy remains nearly 100\% in the training dataset but gradually deteriorates in the testing dataset. Conversely, as SNRs increase, baseline schemes show gradual improvement in authentication accuracy. However, regardless of SNR levels, the proposed scheme consistently outperforms baseline methods due to its consideration of CSI fingerprint distribution changes caused by user movements, whereas other methods assume independent and identical distribution of CSI fingerprints for each user. At the SNR of 15 dB, the proposed scheme demonstrates an improvement in authentication accuracy ranging from 13.04\% to 36.64\%.

\begin{figure}
\centering
\includegraphics[width=0.35\textwidth]{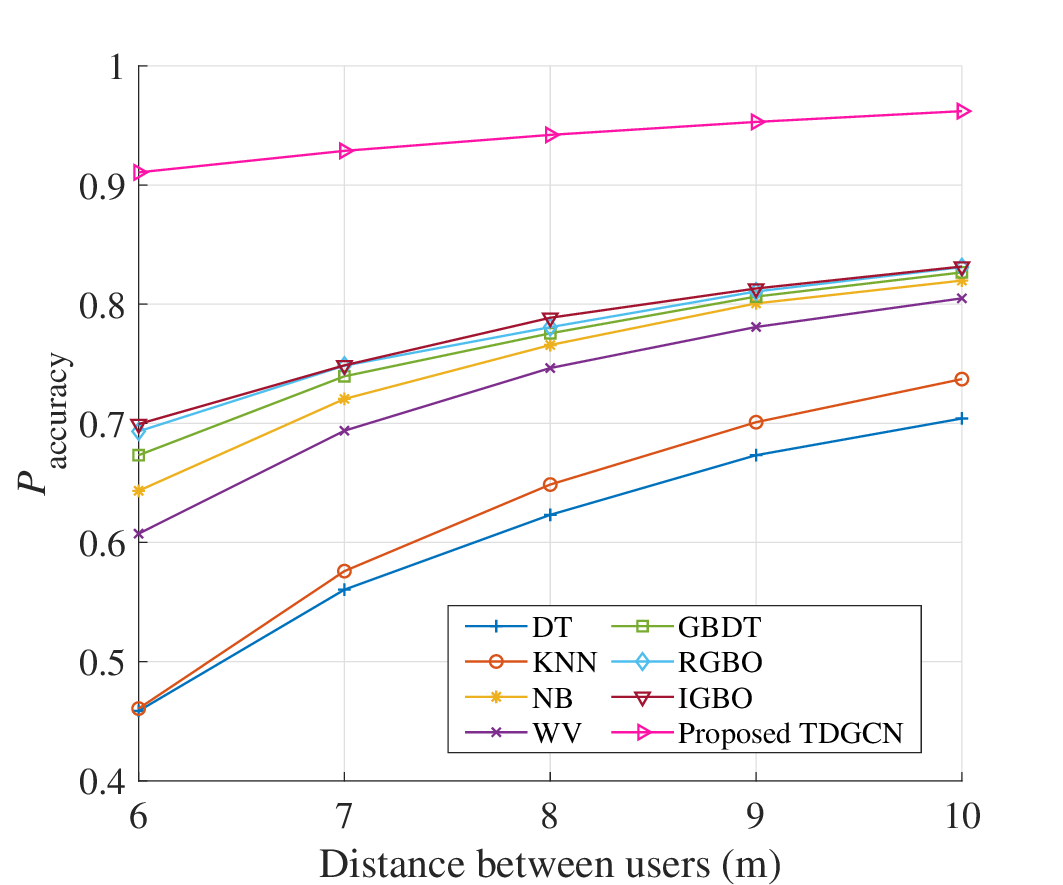}
\caption{Authentication accuracy of different PLA schemes versus different distances between users.}
\label{distance}
\end{figure}

\textit{Performance versus User Distances:}
Fig. \ref{distance} contrasts the authentication accuracy of various PLA schemes against transmitter distances. As the distance between users decreases, the similarity of fingerprints increases, resulting in higher distribution coincidence, thus making it more challenging for the authentication model to differentiate, consequently lowering authentication accuracy. Nevertheless, the proposed scheme consistently outperforms baseline methods by capturing dynamic temporal-spatio features. Fig. \ref{distance} further validates the superiority of the proposed approach.

\begin{figure}
\centering
\includegraphics[width=0.35\textwidth]{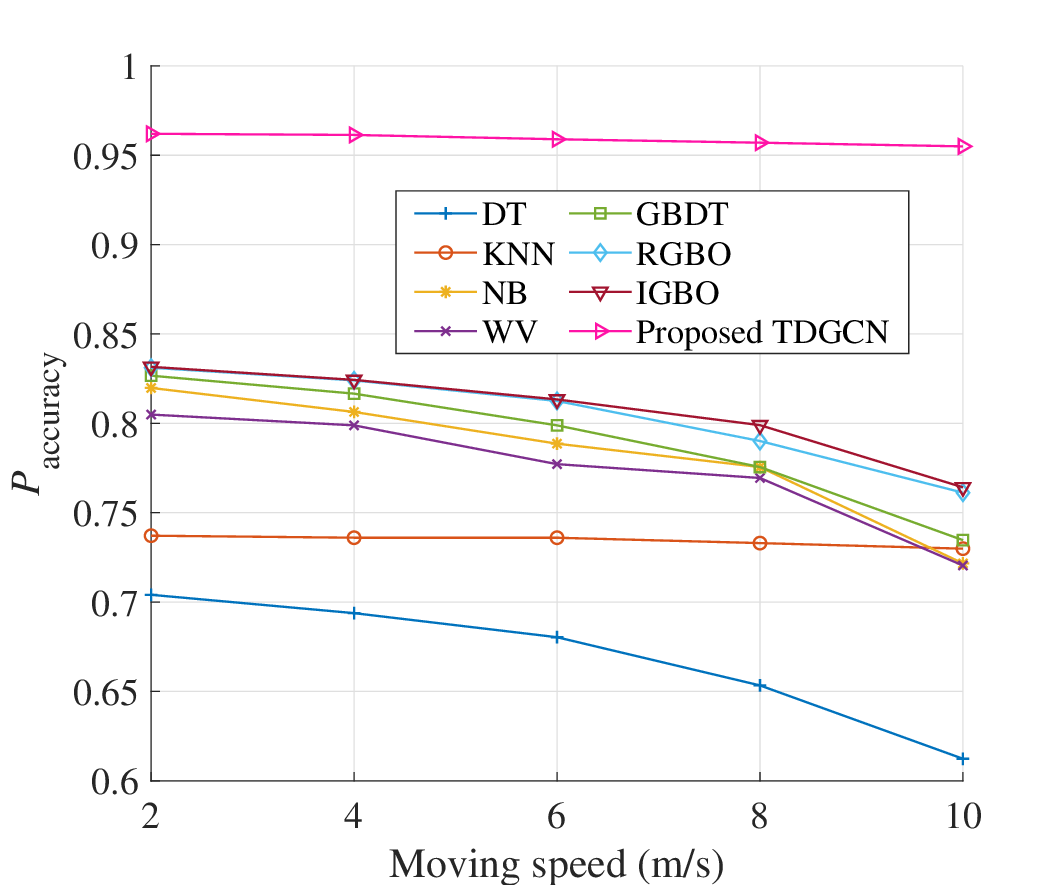}
\caption{Authentication accuracy of different PLA schemes versus different moving speeds of users.}
\label{speed}
\end{figure}

\textit{Performance versus User Speeds:}
Fig. \ref{speed} analyzes the authentication accuracy of different schemes versus user speeds. As users move faster, the distance between adjacent fingerprints increases under the same CSI sampling frequency, leading to lower distribution similarity and decreased performance for most distribution-based authentication models. Although KNN relies on CSI sample distances and is less affected, its feature learning capability is limited, resulting in significantly lower authentication accuracy compared to the proposed TDGCN-based scheme.



\section{Conclusions}
This paper introduces a TDGCN-based PLA scheme, aimed at identifying mobile multi-users in EI-aided IIoT. Leveraging TCNs and dynamic GNNs, the model learns the temporal evolution of each CSI dimension feature and the spatio-temporal dynamics between CSI sequences. Dynamic GINs and cascade pooling mechanisms are utilized to retain learned information while mitigating computational complexity. Simulation results confirm the efficacy of the proposed scheme.
\bibliography{ref.bib}
\bibliographystyle{IEEEtran}

\vfill

\end{document}